\documentclass[%
 superscriptaddress,
 amsmath,amssymb,
 aps,
floatfix,
]{revtex4-2}
\usepackage{graphicx}
\usepackage{dcolumn}
\usepackage{bm}
\newcommand{\RNum}[1]{\uppercase\expandafter{\romannumeral #1\relax}}  
\usepackage[colorlinks=true, linkcolor=cyan, urlcolor=cyan, citecolor=cyan]{hyperref}

\begin{document}

\preprint{APS/123-QED}

\title{Glassy relaxation in de~Vries smectic liquid crystal consisting of bent-core molecules}

\author{Vishnu Deo Mishra}
\affiliation{Soft Condensed Matter Group, Raman Research Institute, C. V. Raman Avenue, Sadashivanagar, Bangalore 560080, India}
\author{G. Pratap}
 \affiliation{Polymer Science and Technology, CSIR-Central Leather Research Institute, Chennai, India}%
\author{Arun Roy}%
 \email{aroy@rri.res.in}
\affiliation{Soft Condensed Matter Group, Raman Research Institute, C. V. Raman Avenue, Sadashivanagar, Bangalore 560080, India}

\date{\today}
\begin{abstract}
We report the experimental investigations on a liquid crystal comprised of thiophene-based achiral bent-core banana shaped molecules. The results reveal the presence of a short range nematic phase at high temperatures and a long-range SmA phase at lower temperatures, which transits to a SmC phase on further cooling the sample. Practically no layer contraction was observed across the SmA to SmC transition, indicating the de Vries nature of the SmA phase. Interestingly, the crystallization does not occur on cooling the sample till 223~K; instead, a glass transition at 271~K was observed. The dielectric spectroscopy studies carried out on the sample reveal the presence of a dielectric mode whose relaxation process is of the Cole-Cole type. The relaxation frequency of the mode was found to drop rapidly with decreasing temperature, further confirming the glassy behavior. The variation of relaxation frequency with temperature follows the Vogel-Fulcher-Tammann equation indicating the fragile glassy nature of the sample.     
\end{abstract}

                              
\maketitle

\section{Introduction}
\begin{figure}[t]
    \centering
    \includegraphics[width=0.5\linewidth]{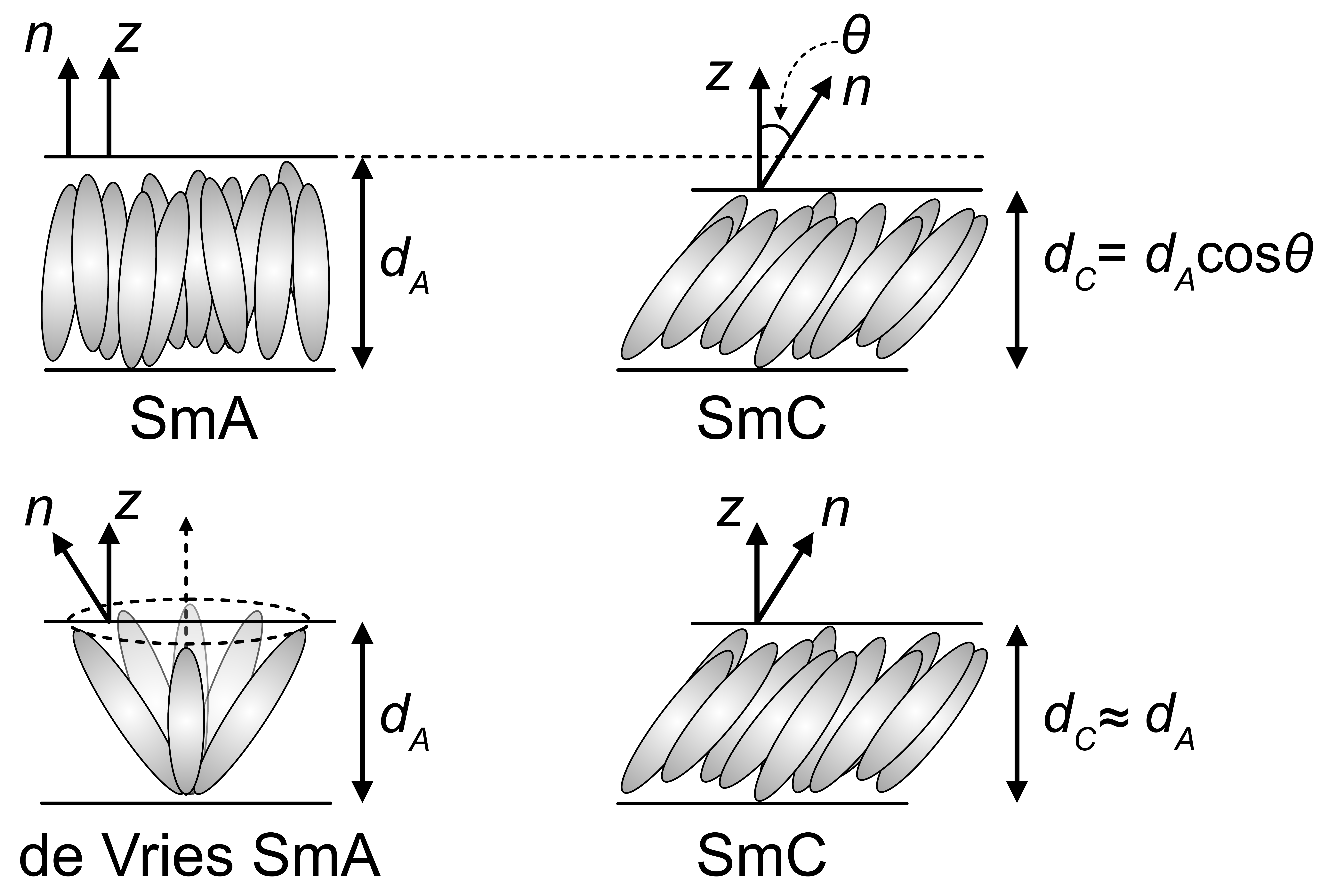}
    \caption{Schematic representation of the molecular arrangement in conventional SmA, SmC, and de Vries SmA phases comprised of calamitic molecules. In the dSmA phase, the molecules are oriented with a large tilt angle with respect to layer normal having long axes randomly distributed on a cone according to the diffuse cone model proposed by Adriaan de Vries  \cite{de_vries_1979_2}.}
    \label{smectic}
\end{figure}

Liquid crystals (LCs) are comprised of molecules with strong shape anisotropy and they exhibit long-range orientational order with or without partial translation order. Thus LCs have the characteristics of crystals while retaining the fluidity of ordinary liquids \cite{Chandrasekhar_1992}. The most commonly observed liquid crystalline phases are known as \textit{nematic} and \textit{smectic} phases. The nematic (N) phase has long range orientational order of the constituent molecules with no translational ordering. The average orientation of the molecular long axes is defined by the unit vector $\hat{n}$ and is known as the director. In the lamellar smectic phases, the anisotropic molecules stack to form layers with translational order along the layer normal $\hat{k}$ but fluid-like order in the layers. The nano-segregation of the electron-rich aromatic core and aliphatic chains of the constituent molecules give rise to such layered structures. In the smectic~A (SmA) phase, the director $\hat{n}$ is parallel to the layer normal $\hat{k}$ with the layer spacing $d$ of the order of molecular length $l$. In the smectic~C (SmC) phase, the director $\hat{n}$ tilts away from the layer normal $\hat{k}$ by a tilt angle $\theta$ with a layer spacing $d$ less than the molecular length $l$. The tilt angle $\theta=\cos^{-1}(d/l)$ is generally found to be temperature dependent in the SmC phase. There is usually a significant reduction of the layer spacing across the transition from the SmA to SmC phase, as depicted in figure~\ref{smectic}. The SmA phase acquires a so called `bookshelf geometry' when confined between parallel bounding plates treated for homogeneous planar alignment of the molecules. In the bookshelf geometry, the constituent molecules align homogeneously on the bounding plates, giving rise to the layers perpendicular to the plates. Across the SmA-SmC transition, the reduction of layer spacing in bookshelf geometry leads to the buckling of layers resulting in the so-called chevron structure \cite{Rieker_1987}. The chevron structure originates due to a mismatch between the smectic layer spacing in bulk and at the surface \cite{Rieker_1987, Limat_1993}. The chevron formation causes various types of defects, such as zig-zag defects, which have been a challenging problem for surface stabilized ferroelectric LC displays \cite{Rieker_1987, Clark_1980, Lagerwall_2006}. This issue can be addressed using materials with no layer contraction across the SmA$-$SmC transition. Such a SmA phase is known as the ``de Vries smectic~A'' phase. Henceforth in our discussion, we shall denote the de Vries SmA phase as `dSmA' phase. Adriaan de Vries first proposed a model structure for this type of SmA phase where molecules in a layer are tilted with respect to the layer normal but with random azimuthal tilt directions. The random distribution of the tilt direction makes this phase optically uniaxial about the layer normal as in the SmA phase. This model structure is termed as diffuse cone model of dSmA phase \cite{de_vries_1979_1, de_vries_1979_2}. Across the transition from the dSmA to the SmC phase, the already tilted molecules in the layers in the dSmA phase choose a particular azimuthal orientation as in the SmC phase with practically no layer contraction as shown in figure~\ref{smectic}. A large number of materials consisting of calamatic (rod-like) molecules have been reported to exhibit the dSmA phase \cite{Lagerwall_2006, Prasad_2009, Roberts_2010, Yoon_2011, Schubert_2017}. But only a few bent-core banana shaped molecules are found to exhibit dSmA phase \cite{Ocak_2011, Sreenilayam_2016, Green_2019, Panarin_2020, Kaur_2022}.

Bent-core (BC) liquid crystals have attracted significant attention in recent years due to their remarkably distinct optical and electro-optical properties arising from the interplay between polarity and chirality \cite{Niori_1996, Link_1997, Jakli_2018, Takezoe_2019}. The bent shape enables the molecules to form a lamellar structure with aligned bending directions, giving rise to layer polarization and thus forming ferro- and antiferroelectric phases \cite{Niori_1996}. Apart from a potential candidate in electro-optic applications due to their fast response time \cite{Takezoe_2019}, the BC molecules exhibit a rich variety of phases, such as orthogonal polar smectic SmAP phase \cite{Sekine_1997, Eremin_2001, Reddy_2011}, polarization modulated structures \cite{Coleman_2003, Vaupotic_2007, Yoon_2010, Chattham_2015}, non polar undulated layer structure \cite{Mishra_2023}, dark conglomerate phase \cite{Hough_2009}, and columnar phases \cite{Takanishi_1999, Pelz_2003, Folcia_2006}.

A large number of studies can be found on glass formation and molecular dynamics studies for calamatic liquid crystals \cite{Zeller_1982, Rzoska_2003, Dierking_2008, Wu_2011, Tarnacka_2013, Delaporte_2016, Deptuch_2022}, but glass forming bent-core molecular system is rarely reported \cite{Rauch_2004}. The formation of the glassy state is another intriguing dynamic slow-down process that retained the interest of scientists for several decades because of their ubiquitous role in technology and diverse applications \cite{Blanshard_1993, Greer_1995, Crowe_1998, Debenedetti_2001}. The glass formation can take place by many routes \cite{Angell_1995}, a conventional one of which is rapid enough cooling of a liquid phase to avoid the nucleation and growth process of crystals. The liquid appears frozen on the time scale of experimental observation. Although rapid cooling is required for glass formation in many liquids, some organic liquids \cite{Dyre_2006}, polymers \cite{Angell_1995}, and liquid crystals \cite{Elschner_1999} can exhibit glass transition at a moderate cooling rate. In the glass-forming liquid crystal, the order can be frozen by preserving the director by external means, such as strong anchoring or applied electric/magnetic field, while cooling the system. This allows the formation of partially ordered glasses retaining all the qualitative features of the liquid crystal phase, which could be exploited in possible applications such as wave plates, holography, and optical storage \cite{Dierking_2008}. Recently it has been pointed out that for some materials, a smectic transition from high temperature nematic phase can be frustrated or completely circumvented to produce glasses with tunable liquid crystalline order that can be utilized in organic electronic applications \cite{Teerakapibal_2018, Chen_2020}. 

In this paper, we investigate the physical properties of liquid crystalline phases of an achiral thiophene-based bent-core liquid crystal using various experimental techniques. The compound exhibits a high temperature short range nematic phase followed by two long range smectic phases. We report two key findings from our experimental studies. Firstly, we find that the higher temperature smectic phase is the dSmA phase which transits to the SmC phase on decreasing temperature with practically no layer contraction. Unlike the typical BC liquid crystals, both the smectic phases are non-polar with calamatic type phase behavior. Secondly, both the smectic phases show a dielectric relaxation mode whose relaxation frequency decreases sharply on lowering the temperature suggesting a glassy behavior. The dielectric relaxation of the mode is of Cole-Cole type, and the temperature variation of relaxation frequency follows the empirical Vogel-Fulcher-Tammann (VFT) equation. This suggests the \textit{fragile} nature of the glass for our sample with fragility parameter $D \approx$ 3 \cite{Angell_1995}. The calorimetric studies further confirmed the glass transition at a temperature of about 271~K, which qualitatively agrees with the dielectric studies. We propose a simple model structure for the dSmA phase to account for the experimental observations. To the best of our knowledge, this is the first report of a bent-core liquid crystal exhibiting a dSmA phase and also showing glassy behavior.

\section{Experimental}\label{Experimental}
Polarized Optical Microscopy (POM) investigations of the sample were conducted using Olympus BX~50 microscope equipped with a hot stage (Linkam LTS420E) and a digital camera (Canon EOS~80D). For homeotropic alignment of the molecules, a thin sample was sandwiched between a clean glass slide and a glass cover slip. In addition, commercially available LC cells from (INSTEC Inc.) were employed for the planar and homeotropic alignment of the sample. The LC cells contain indium tin oxide (ITO) coated glass plates serving as an electrode for electro-optic and dielectric studies. The sample was introduced into the LC cell by capillary action in its isotropic phase using a hot plate. The electric polarization of the sample in the LC phases was investigated using a triangular wave voltage technique \cite{Miyasato_1983} with varying amplitudes and frequencies for planar as well as homeotropically aligned samples. The current response of the sample was measured by monitoring the voltage drop across a 1~k$\Omega$ resistor connected in series with the sample cell.

To evaluate the effective birefringence of a planar aligned sample, we measured the variation of average intensity transmitted through the sample as a function of temperature. The planar aligned sample was kept on the microscope stage at an angle of maximum transmittance between crossed polarizers, and images were taken with varying temperatures by introducing a red filter of wavelength 700~nm in the light path. The average transmitted intensity was computed from the POM images using MATLAB. 

The steady state field induced electro-optical response of a planar aligned sample was measured using a He-Ne laser with the application of a triangular AC voltage. The sample was kept between crossed polarizers at an orientation of maximum transmittance, and the transmitted intensity was recorded using a low-noise high gain photodiode connected to a mixed signal oscilloscope (Agilent Technologies MSO6012A). The intensity was normalized using $I_{nor} = (I-\bar{I})/\bar{I}$, where $I$ is the measured transmitted intensity through the sample and $\bar{I}$ is the mean value of the transmitted intensity. 

Variable temperature x-ray diffraction (XRD) measurements were conducted using a DY~1042-Empyrean (PANalytical) diffractometer with CuK$_\alpha$ radiation of wavelength 1.54~\AA{} and a PIXcel 3D detector. The samples were filled in Lindemann capillary tubes with an outer diameter of 1~mm. The XRD intensity profiles were measured at different temperatures on cooling the sample from its isotropic phase.

Dielectric measurements were performed on the sample filled in commercially available LC cell using a high-performance impedance analyzer (Novocontrol Alpha-A) in the frequency range of 1~Hz to 10~MHz. The measurements were limited to 1 Hz due to the dominating charge current contribution at lower frequencies. A sinusoidal AC voltage with an rms amplitude of 0.5~Volts was used during the dielectric measurement without applying a bias voltage. The temperature of the sample was monitored using a homemade temperature controller with temperature stability of 0.1~K. To study the variation of the dielectric constant of the sample in the high temperature range, a complimentary homemade setup with a frequency range of 1~Hz to 100~kHz was also employed. The temperature of the sample was monitored by a microscope hot stage equipped with a temperature controller (INSTEC Inc.). In this setup, a sinusoidal AC voltage of rms amplitude 0.5~Volts was applied to the sample cell and a resistance of 1~k$\Omega$ connected in series. A lock-in amplifier (Stanford Research SR830) was utilized to measure the amplitude and phase of the voltage drop across the 1~k$\Omega$ resistance. The impedance analysis was used to determine the capacitance of the LC cell with and without the sample. The ratio of these measured capacitances gives the real part of the effective dielectric constant of the sample. 

\begin{figure}[!t]
    \centering
    \includegraphics[width=0.5\linewidth]{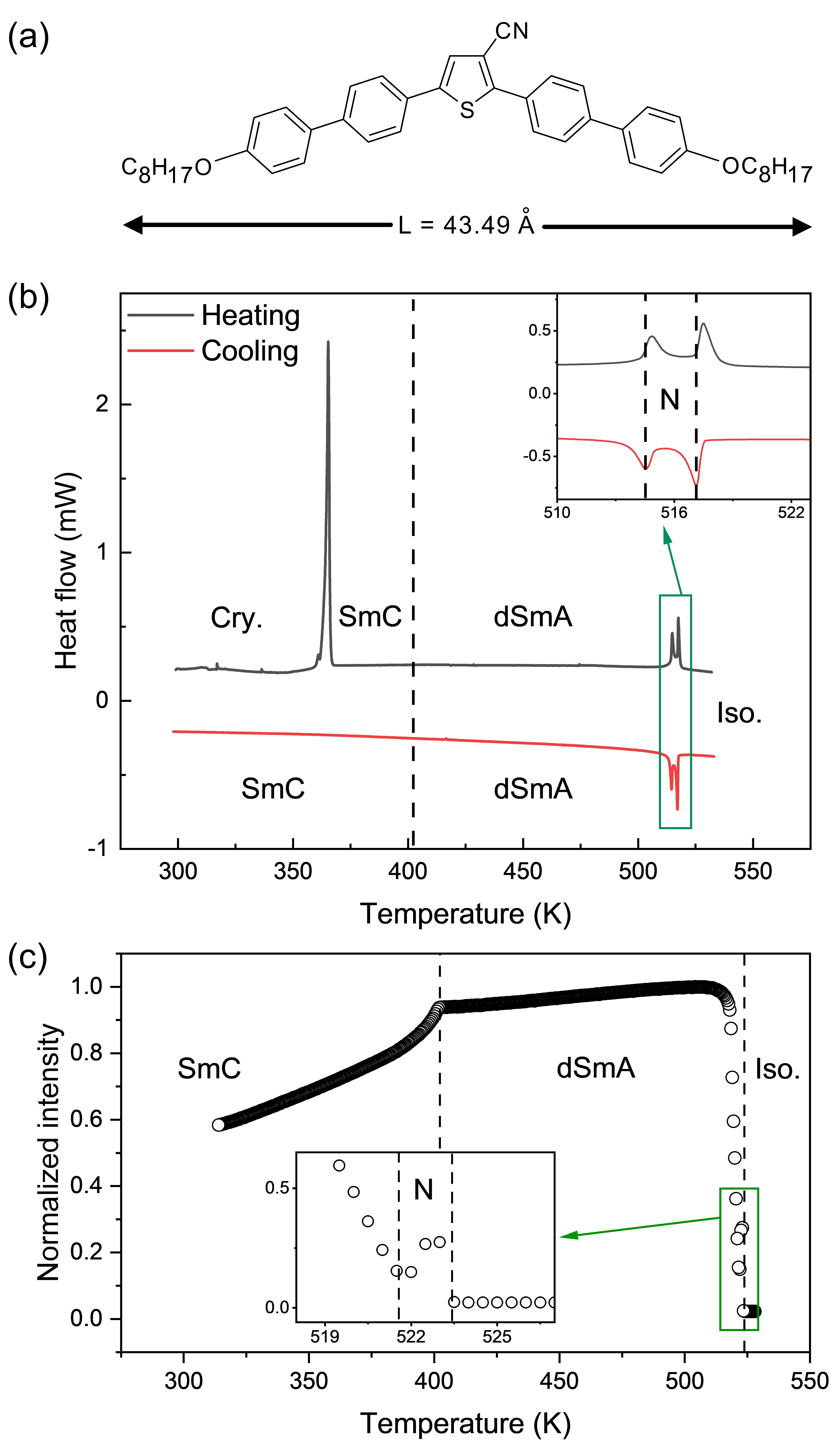}
    \caption{(a) The molecular structure of the compound BTCN8. (b) DSC thermogram of the compound BTCN8 with heating and cooling rate of 5~K/min. The inset demonstrates the existence of a small range nematic phase. The sample does not crystallize on cooling till room temperature. The mesomorphic properties of the sample were retained for several weeks, and the melting transition was observed only on the first heating. (c) The temperature variation of the optical transmittance through a planar aligned sample of thickness 5~$\mu$m kept between crossed polarizers while cooling from the isotropic phase. The optical transmittance clearly detects the following phase transition sequence: Isotropic (523.5~K) N (521.5~K) dSmA (402~K) SmC. The inset reveals the existence of a nematic phase with a small range of temperature. All the observed phases are enantiotropic.}
    \label{dsc}
\end{figure}

\section{Results and discussion}
\subsection{Phase sequence}
The sample used for our experimental studies is \emph{2,5-bis(4$'$-(octyloxy)-[1,1$'$-biphenyl]-4-yl)thiophene-3-carbonitrile} and is denoted as BTCN8. The chemical characterization and preliminary studies on this compound have already been published \cite{Pratap_2019}. The molecule has a bent-core banana shape with a central thiophene ring, as depicted in Figure~\ref{dsc}a. The opening angle and the molecular length, calculated from the energy minimized molecular structure, are about 143$^\circ$ and 43.5~\AA{}, respectively. A strongly polar carbonitrile group attached to the central thiophene ring enhances the net dipole moment of the molecule at an angle to the long axis \cite{Pratap_2019}.  

Differential Scanning Calorimetry (DSC) studies were performed on the sample to detect the various phase transitions in heating and cooling cycles at a rate of 5~K/min. The DSC thermograms of the sample are shown in figure~\ref{dsc}b. The DSC thermogram reveals that the sample melts while heating at 365~K, which on further heating, shows two transition peaks at 515.9~K and 518.6~K and goes to the isotropic phase. On cooling from the isotropic phase, two transitions were observed at 517.4~K and 514.9~K, and no other transitions were observed on further cooling the sample till room temperature. The inset in figure~\ref{dsc}b shows the expanded view of the two closely spaced transitions. Thus, the DSC thermogram clearly indicates the existence of two mesophases for the compound BTCN8 below its isotropic phase. However, detailed polarizing optical microscopy (POM) studies detect another transition at 402~K while cooling the sample. In POM studies, the optical transmittance through a planar aligned 5~$\mu$m thick sample kept between crossed polarizers was measured while cooling from the isotropic phase. The rubbing direction of the sample cell was kept at an angle of 45$^\circ$ from the polarizer direction for maximum transmittance. Figure~\ref{dsc}c displays the normalized transmitted intensity as a function of temperature, which clearly detects all the phase transitions. The inset of figure~\ref{dsc}c shows the existence of the short-range nematic phase. The discontinuous change in the slope of the transmitted intensity curve at 402~K corresponds to an additional phase transition. Based on the experimental studies, we identify the following phase sequence for the compound BTCN8 on cooling till room temperature: \[\text{Isotropic $\xrightarrow{517.4~\text{K}}$ N $\xrightarrow{514.9~\text{K}}$ de Vries SmA $\xrightarrow{402~\text{K}}$ SmC}\]

All the observed mesophases are found to be enantiotropic. The absence of a DSC peak across the dSmA to SmC phase transition indicates the second-order nature of this transition. The SmC phase became increasingly viscous on lowering the temperature while retaining the fluidity till room temperature. The sample did not crystallize on cooling. Instead, a glass transition was observed at 271~K, which will be discussed later. 

\subsection{X-ray diffraction studies}
The variable temperature x-ray diffraction (XRD) studies were carried out at various temperatures to investigate the molecular organization in the observed liquid crystalline phases. 
\begin{figure}[!t]
    \centering
    \includegraphics[width=0.5\linewidth]{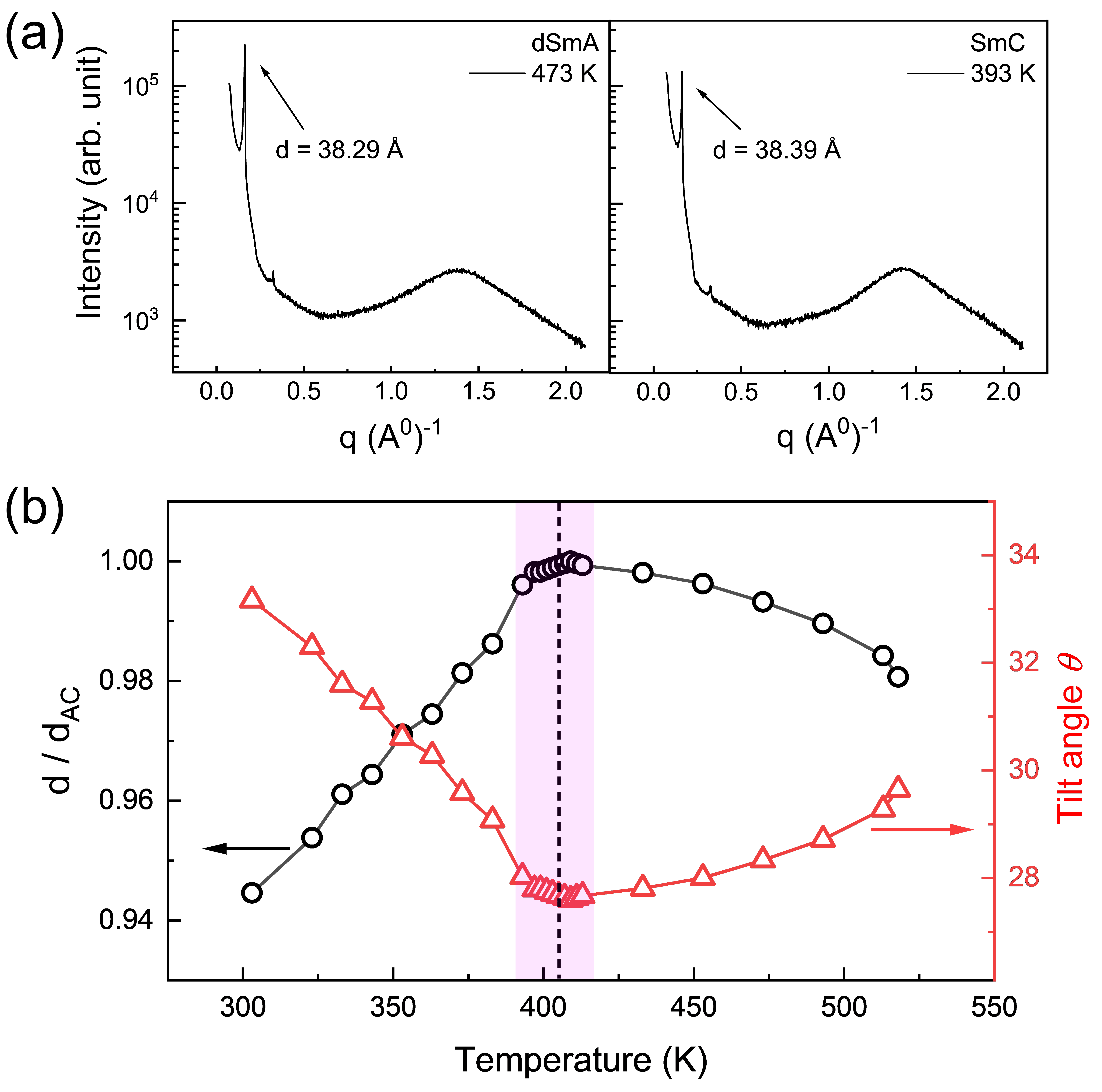}
    \caption{(a) The XRD intensity profile for the compound BTCN8 in dSmA and SmC phases at temperatures 473~K and 393~K, respectively. (b) The temperature variation of the ratio $d/d_{AC}$ and tilt angle $\theta$; which were obtained from the XRD data and estimated molecular length. The layer thickness remains almost unchanged across the dSmA-SmC transition showing the de Vries nature of the SmA phase. The vertical dotted line indicates the dSmA to the SmC transition temperature. Across the transition depicted by the pink shaded region, the layer contraction is only about 0.07\%.}
    \label{xrd}
\end{figure}
In the highest temperature short range nematic phase, a single diffuse peak was observed in the wide angle region with no sharp peak at the small angle region, which is characteristic of the Nematic phase. In the dSmA and SmC phases, the XRD intensity profiles as a function of the scattering vector $q$ are shown in figure~\ref{xrd}a. In both of these phases, two sharp peaks were observed in the small angle region at $q$ values in the ratio of 1:2. In addition, a diffused broad peak was observed in the wide angle region centered about $q = 1.36$~\AA{}$^{-1}$ and $q = 1.43$~\AA{}$^{-1}$ in the dSmA and SmC phases, respectively. The shifting of the maximum of the wide angle diffuse peak towards a higher value of $q$ on cooling is due to increased molecular packing density. The XRD results indicate a lamellar molecular organization with liquid-like order within the layers. The layer spacings calculated from the XRD data in the dSmA and SmC phases are 38.39~\AA{} and 38.29~\AA{}, respectively, which do not vary appreciably with temperature. The molecular length $l$ of the compound BTCN8 is 43.5~\AA{}, which is significantly larger than the observed layer spacing $d$ in both the smectic phases. This indicates that the molecules are tilted within the layers. The tilt angle with respect to the layer normal can be estimated using $\theta=\cos^{-1}(d/l)$. The temperature variation of the tilt angle $\theta$ is shown in figure~\ref{xrd}b. The tilt angle $\theta$ varies slightly across the whole temperature range and attains a shallow minimum value of about 28$^\circ$ near the dSmA to SmC transition. Figure~\ref{xrd}b also shows the temperature variation of the normalized layer spacing $d/d_{AC}$, where $d_{AC}$ is the maximum value of layer spacing in the dSmA phase close to the transition. It is clear from figure~\ref{xrd}b that the layer spacing does not change appreciably across the transition from the dSmA to the SmC phase. The layer contraction in the SmC phase at 10~K below the dSmA-SmC transition temperature is only about 0.07\%. These observations confirm that the higher temperature smectic phase is a de Vries-type SmA phase. In the dSmA phase, the molecules are tilted within the layers having their tilt directions distributed uniformly in the azimuthal plane giving rise to optically uniaxial texture about the layer normal. In contrast to the conventional SmA phase, the layer spacing in the dSmA phase is significantly less than the molecular length. The large opening angle of the bent structure of the BTCN8 molecules and a strong dipole moment due to the polar cyano group projecting in the lateral direction perhaps favor the tilted non-polar molecular organization in the layers \cite{Govind_2001}.

Across the transition from the dSmA to the SmC phase, the random directions of the already tilted molecules get correlated, giving rise to a uniformly tilted configuration with no considerable change in the layer thickness. It can be seen from figure~\ref{xrd}b that the layer spacing is maximum near the dSmA-SmC transition and decrease slightly from this value in the dSmA phase at higher temperatures and in the SmC phase at lower temperatures. The increase in the layer thickness in the dSmA phase with decreasing temperature can be attributed to the stretching of the alkyl chain with increasing packing density. This trend of negative thermal expansion of the layer spacing in the dSmA phase has also been observed in other materials exhibiting de Vries-type SmA phases \cite{Roberts_2010, Schubert_2017}. In our sample, a slight decrease in layer spacing was observed in the SmC phase at lower temperatures. However, the observed layer contraction in the whole temperature range of about 225~K in the dSmA and the SmC phases is only about 5\%. The POM studies discussed in the following section further confirmed the de Vries nature of the SmA phase.


\subsection{POM and electro-optic measurements}

\begin{figure}[!t]
    \centering
    \includegraphics[width=0.5\linewidth]{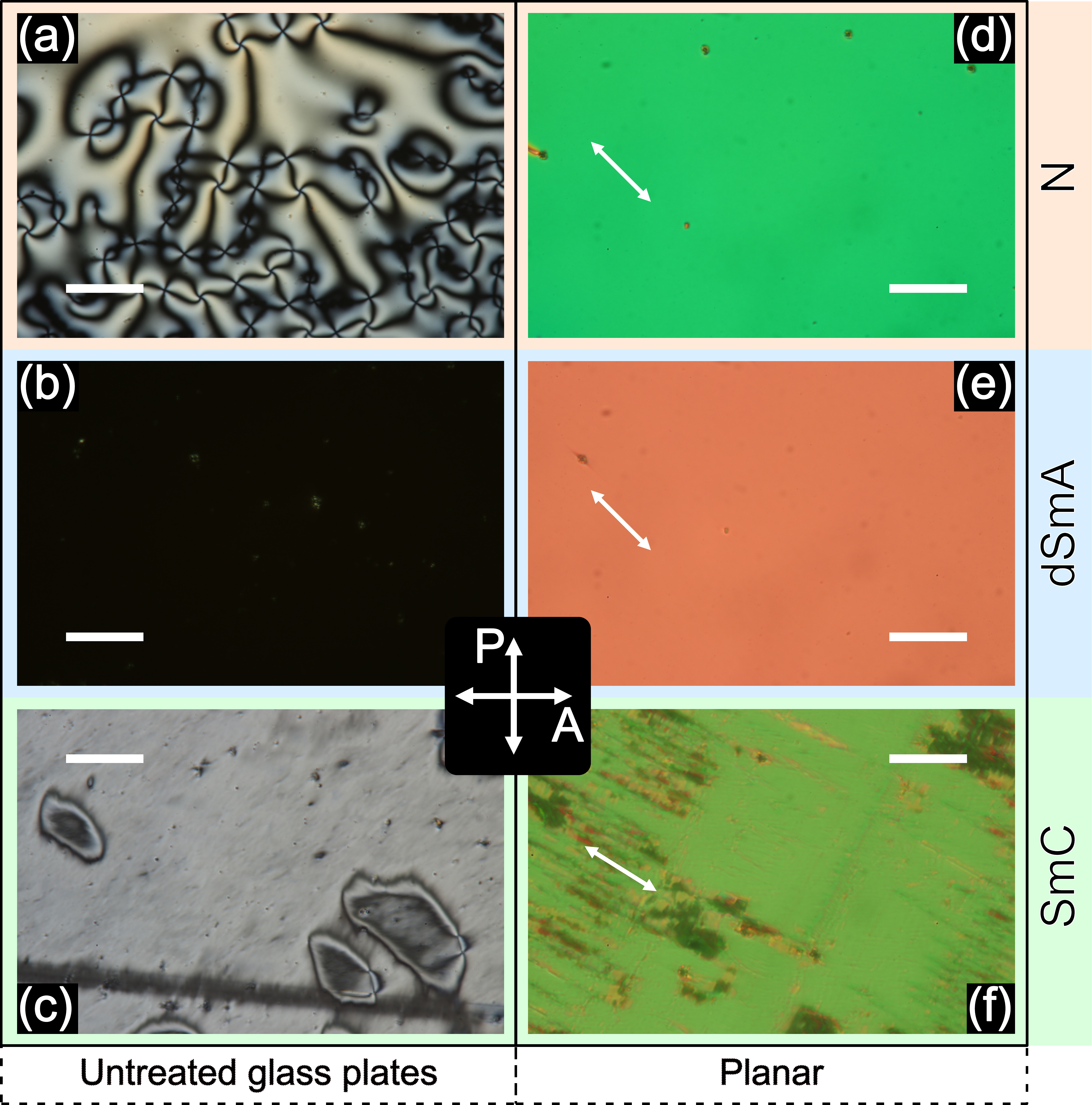}
    \caption{POM texture of homeotropically aligned thin sample kept between a clean glass plate and a cover slip at (a) 522~K, (b) 518~K, and (c) 388~K; and that of the planar aligned sample of thickness 5~$\mu$m at (d) 522~K, (e) 508~K, and (f) 333~K. The POM textures were taken under crossed polarizers conditions while cooling the samples from their isotropic phase. \textbf{\textit{R}} indicates the rubbing direction, and the crossed arrows denote the positions of the polarizers. The scale bar denotes a length of 50~$\mu$m.}
    \label{pom}
\end{figure}

The POM observations were carried out on thin samples to further characterize the different phases. The POM textures were observed between crossed polarizers while cooling the sample from the isotropic phase. A thin sample sandwiched between a clean glass plate and a cover slip tends to align homeotropically. The nematic phase below the isotropic phase exhibits a characteristic schlieren texture as shown in figure~\ref{pom}a. Upon cooling to the dSmA phase, the molecules align homeotropically, and the texture appears completely dark, as shown in figure~\ref{pom}b. The dark texture remains invariant on rotating the sample on the microscope stage, confirming the uniaxial nature of the dSmA phase. This observation and the intralayer molecular tilt observed in the XRD studies discussed earlier confirm the de Vries nature of the SmA phase. On further cooling, a birefringent schlieren texture was observed in the lower temperature SmC phase as shown in figure~\ref{pom}c. The birefringent schlieren texture in this homeotropic geometry and the tilted molecular organization in the layers observed in XRD studies suggest the tilted smectic order. Only unit strength defects were observed in the schlieren texture of this SmC phase, indicating the synclinic organization of the molecules in the layers.

In a planar aligned LC cell, the sample aligns homogeneously along the rubbing direction in the nematic and dSmA phase. The textures in the nematic and the dSmA phase are brightest when the rubbing direction is kept at 45$^\circ$ with respect to the polarizer, as shown in figure~\ref{pom}d and \ref{pom}e, respectively. In the dSmA phase, the layers are perpendicular to the glass plates adopting a so-called bookshelf geometry with an optic axis along the layer normal. The smooth texture of the dSmA phase breaks down into two types of domains as shown in figure~\ref{pom}f in the lower temperature SmC phase. The degeneracy in the azimuthal tilt orientations of the molecules in the dSmA phase is lifted during the transition to the SmC phase. It leads to the formation of two surface stabilized regions with symmetrically opposite optical tilt, as shown in figure~S1 of Supplementary Information (SI).

\begin{figure}[t]
    \centering
    \includegraphics[width=0.5\linewidth]{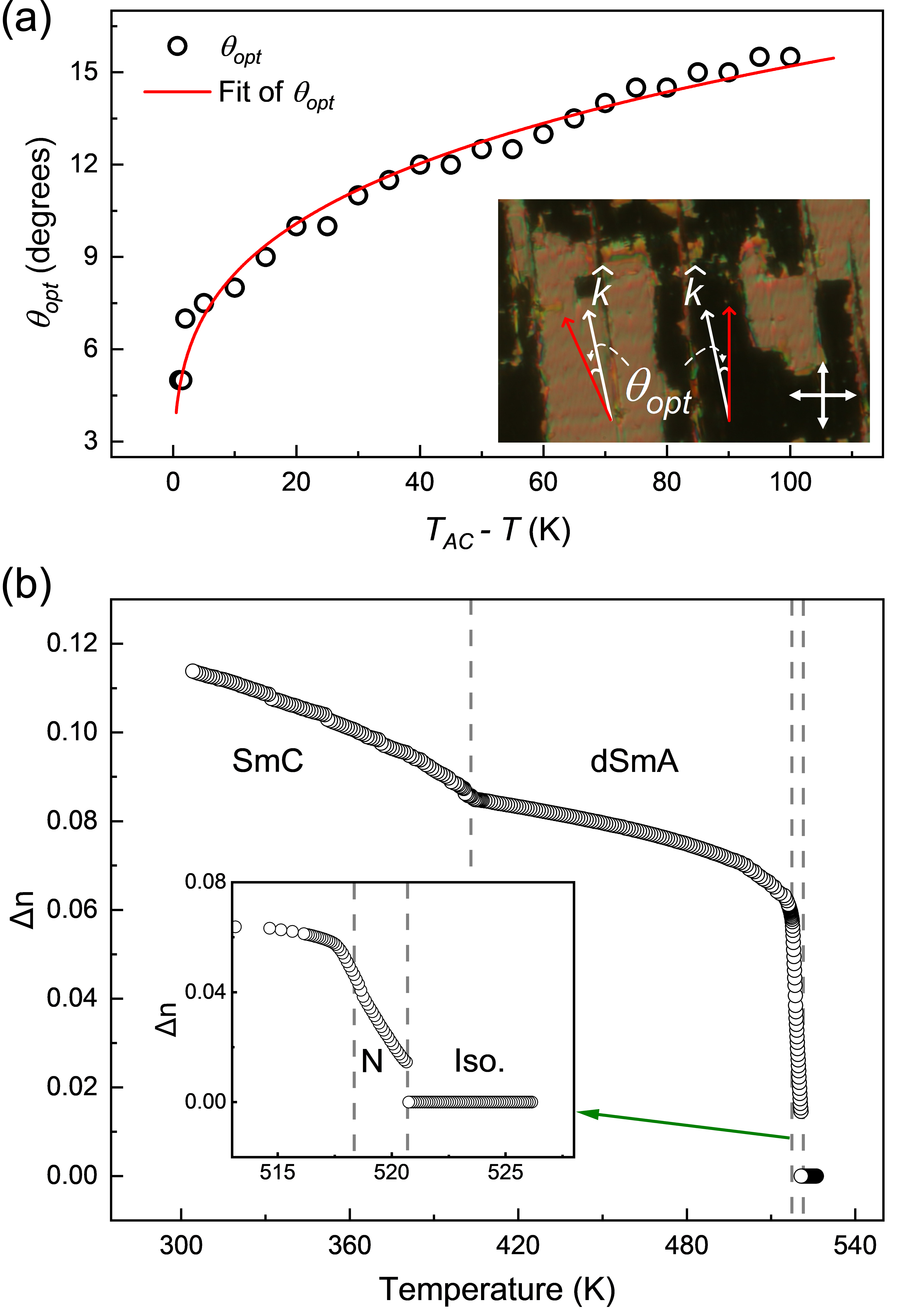}
    \caption{(a) The variation of optical tilt angle $\theta_{opt}$ as a function of ($T_{AC}-T$) in the SmC phase. The solid line shows the fit to the experimental data using eqn. \ref{OP}. The inset of (a) shows the oppositely tilted domains in the SmC phase with the layer normal denoted by unit vector $\hat{k}$. In both the domains, $\hat{k}$ are parallel, but the optic axes are tilted on the opposite side of $\hat{k}$ giving rise to the optical contrast between crossed polarizers. The white and red arrows represent the direction of the layer normal $\hat{k}$ and the optic axis, respectively, in the two domains. (b) Temperature variation of the effective birefringence ($\Delta n$) of a planar aligned sample while cooling from the isotropic phase. The inset of (b) shows the magnified view of the data in the short temperature range of the nematic phase.}
    \label{pom1}
\end{figure}

Though the layer thickness and hence the molecular tilt does not change appreciably across the dSmA to SmC transition, the increasing correlation of the molecular tilt direction in the SmC phase gives rise to an optic axis away from the layer normal. The angle between the optic axis and the layer normal is defined as the optical tilt angle $\theta_{opt}$. The $\theta_{opt}$ of a given domain of a 5~$\mu$m thick planar aligned sample in the SmC phase was measured using POM as a function of temperature. Figure~\ref{pom1}a shows the variation of $\theta_{opt}$ as a function of $T_{AC}-T$, where $T_{AC}$ is the dSmA to SmC transition temperature and $T$ is the measured temperature. The inset of figure~\ref{pom1}a depicts the POM texture of two opposite tilted domains under crossed polarizers at 287~K. The rotation angle between dark states in domains of opposite tilt orientations is 2$\theta_{opt}$. As can be seen from the figure~\ref{pom1}a, the value of $\theta_{opt}$ increases continuously from zero and tend to saturate at lower temperatures. Thus, $\theta_{opt}$ can be treated as an order parameter for the dSmA to SmC transition, which is expected to vary as  
\begin{equation}
\theta_{opt}(T)=\theta_0(T_{AC}-T)^\beta
\label{OP}
\end{equation}
in the SmC phase according to the generalized Landau theory of phase transition. Here, $\beta$ is the tilt exponent related to the nature of the phase transition. The generalized Landau theory predicts $\beta=0.5$ for second-order transition, whereas $\beta=0.25$ for a tricritical point corresponding to crossover from second- to first-order transition. The solid line in figure~\ref{pom1}a shows the fit to the experimental data using eqn.~\ref{OP} with exponent $\beta=0.255 \pm 0.009$ for our sample. The obtained $\beta$ value suggests that the dSmA to SmC transition in our sample is close to the tricritical point. The absence of a DSC peak associated with this transition in our sample suggests that the transition is of second order in nature but close to tricriticality. Similar exponent values have also been reported for other materials exhibiting dSmA to SmC phase transition \cite{Song_2013, Singh_2016}.

We also measured the effective birefringence of a planar aligned sample of thickness 5~$\mu$m as a function of temperature in the liquid crystalline phases. The birefringence was measured by monitoring the transmitted intensity through the sample between crossed polarizers using POM. The transmitted intensity through the sample can be written as, 
\begin{equation}
I = \frac{I_0}{2}\sin^2(2\Psi)(1-\cos\Delta\Phi)
\label{delta_n}
\end{equation}
where $I_0$ is the intensity of the incident light, $\Psi$ is the angle between the local optic axis and the polarizer, and $\Delta\Phi = (2\pi\Delta nd)/\lambda$ is the phase difference introduced by the sample between the ordinary and extraordinary rays. Here, $d$ is the sample thickness, $\lambda$ is the wavelength of the incident light, and $\Delta n$ is the effective birefringence of the sample. In our experiments, the angle $\Psi$ was set to 45$^\circ$ for maximum transmittance. The details of the experimental method to measure the transmitted intensity are described in section~\ref{Experimental}, and the effective birefringence $\Delta n$ was calculated using eqn.~\ref{delta_n}. The variation of $\Delta n$ as a function of temperature is shown in figure~\ref{pom1}b. While cooling from the isotropic phase, a discontinuous jump in the $\Delta n$ was observed at the Isotropic-N transition temperature. The value of $\Delta n$ continued to increase in the N phase, and a slope change at the N-dSmA transition temperature was observed. The $\Delta n$ value increases slightly with decreasing temperature and tends to saturate at lower temperatures in the dSmA phase. The $\Delta n$ again started to increase with decreasing temperature from the dSmA to SmC transition temperature. The low value of the birefringence in the dSmA phase ($\Delta n \approx $ 0.08) compared to that of the SmC phase supports the de Vries nature of the SmA phase \cite{Lagerwall_2006}.

The spontaneous electric polarization of the sample in the dSmA and SmC phase was investigated using a triangular wave voltage technique \cite{Miyasato_1983}. A voltage of amplitude 50~V was applied across the planar aligned sample of thickness 5~$\mu$m. The current responses of the sample are shown in figure~S2 of SI. No current peak associated with polarization reversal was observed, indicating that the layers in both the smectic phases do not possess spontaneous polarization.

Despite the absence of polarization, a clear reversible field-induced change in the texture was observed in the SmC phase for a planar aligned sample (see figure~S3 of SI). The electro-optical response of the planar aligned sample kept between crossed polarizers was measured in the SmC phase under the application of a triangular wave voltage. The observed optical response was found to be at twice the frequency of applied voltage (see figure~S4 of SI). This observation also suggests the absence of polarization in the layers. The observed electro-optical response arises due to the quadratic coupling between the applied electric field and the dielectric anisotropy of the sample. No such optical response was found in the dSmA phase with an applied field as high as 20~V/$\mu$m. 

\subsection{Dielectric studies}
\begin{figure}[t]
    \centering
    \includegraphics[width=0.5\linewidth]{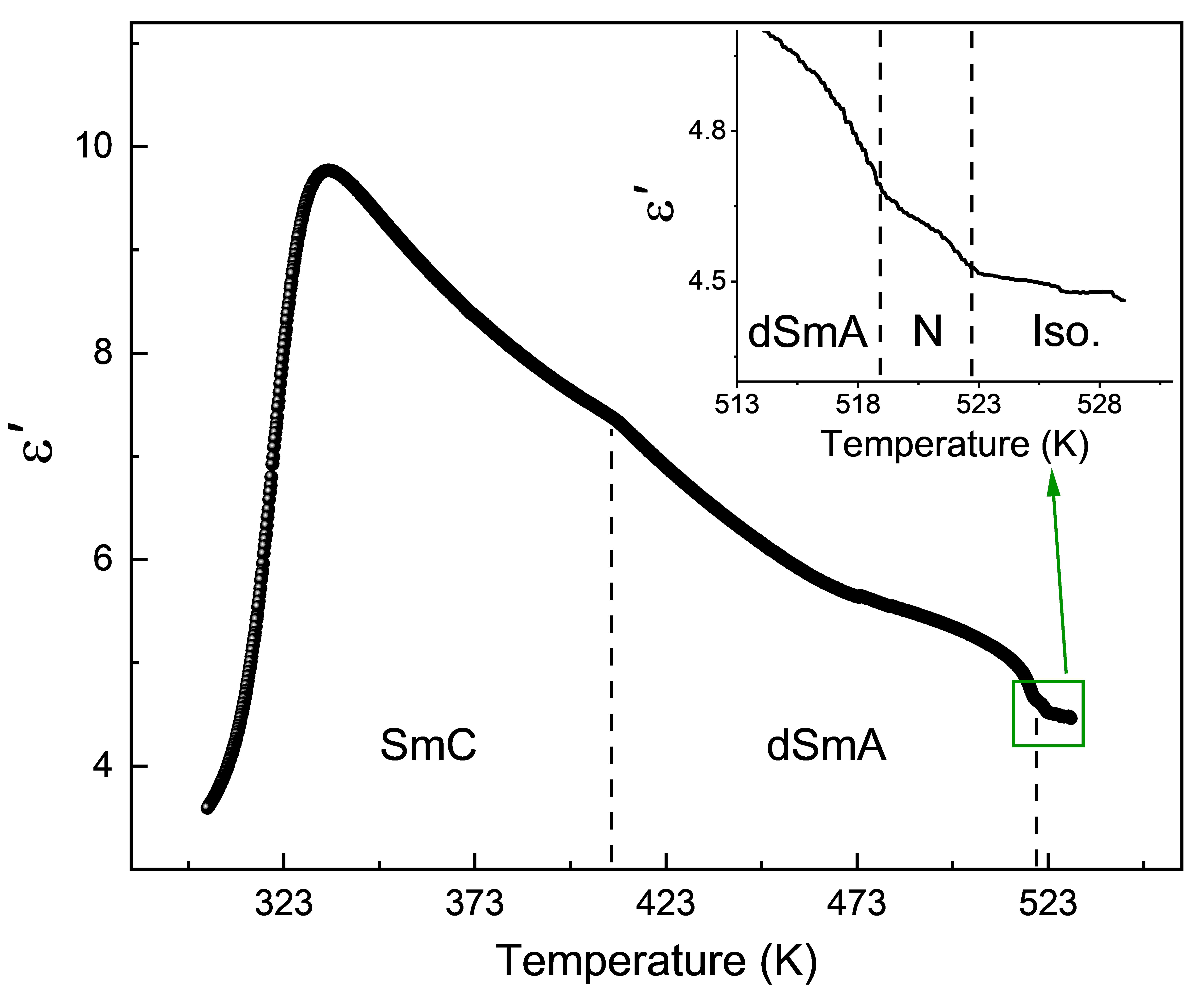}
    \caption{The variation of the effective dielectric constant of the compound BTCN8 in a planar aligned LC cell of sample thickness 5 $\mu$m as a function of temperature. The data clearly detects the various phase transitions and are in agreement with the DSC and POM observations. The dielectric constant's low value indicates the absence of spontaneous polarization in the sample. The inset shows the magnified view of the marked region.}
    \label{dielectric}
\end{figure}

The dielectric properties of the samples were investigated using the LC cells for planar alignment. The methods used for the dielectric measurements are described in section~\ref{Experimental}. The real part of the dielectric constant ($\epsilon^\prime$) was measured by applying a sinusoidal AC voltage of rms amplitude 0.5~V and frequency 5641~Hz while cooling the sample from its isotropic phase. Figure~\ref{dielectric} shows the temperature variation of $\epsilon^\prime$ in the observed liquid crystalline phase of the sample, which clearly detects the transitions between the phases. The observed transition temperatures agree with the DSC and POM measurements. The increase of $\epsilon^\prime$ in the planar aligned nematic phase while cooling from the isotropic phase suggests negative dielectric anisotropy of the sample. The $\epsilon^\prime$ continued to increase on further decreasing the temperature till 333~K and sharply decreased below it. This sharp decrease does not correspond to a phase transition. Rather, it occurs due to dielectric relaxation, as discussed later.  

To study the dielectric relaxation behavior of the samples, the dielectric permittivity was measured as a function of frequency in the range of 1~Hz to 10~MHz at different temperatures for planar aligned samples. The frequency dependent complex dielectric permittivity of the sample can be written as,
\begin{equation}
\epsilon^{*}(f) = \epsilon^{\prime}(f) + i\epsilon^{\prime\prime}(f)
\end{equation}
where $\epsilon^{\prime}$ and $\epsilon^{\prime\prime}$ are real and imaginary parts of the complex dielectric permittivity, respectively, and $f$ being the frequency of the applied field.

\begin{figure}[t]
    \centering
    \includegraphics[width=0.5\linewidth]{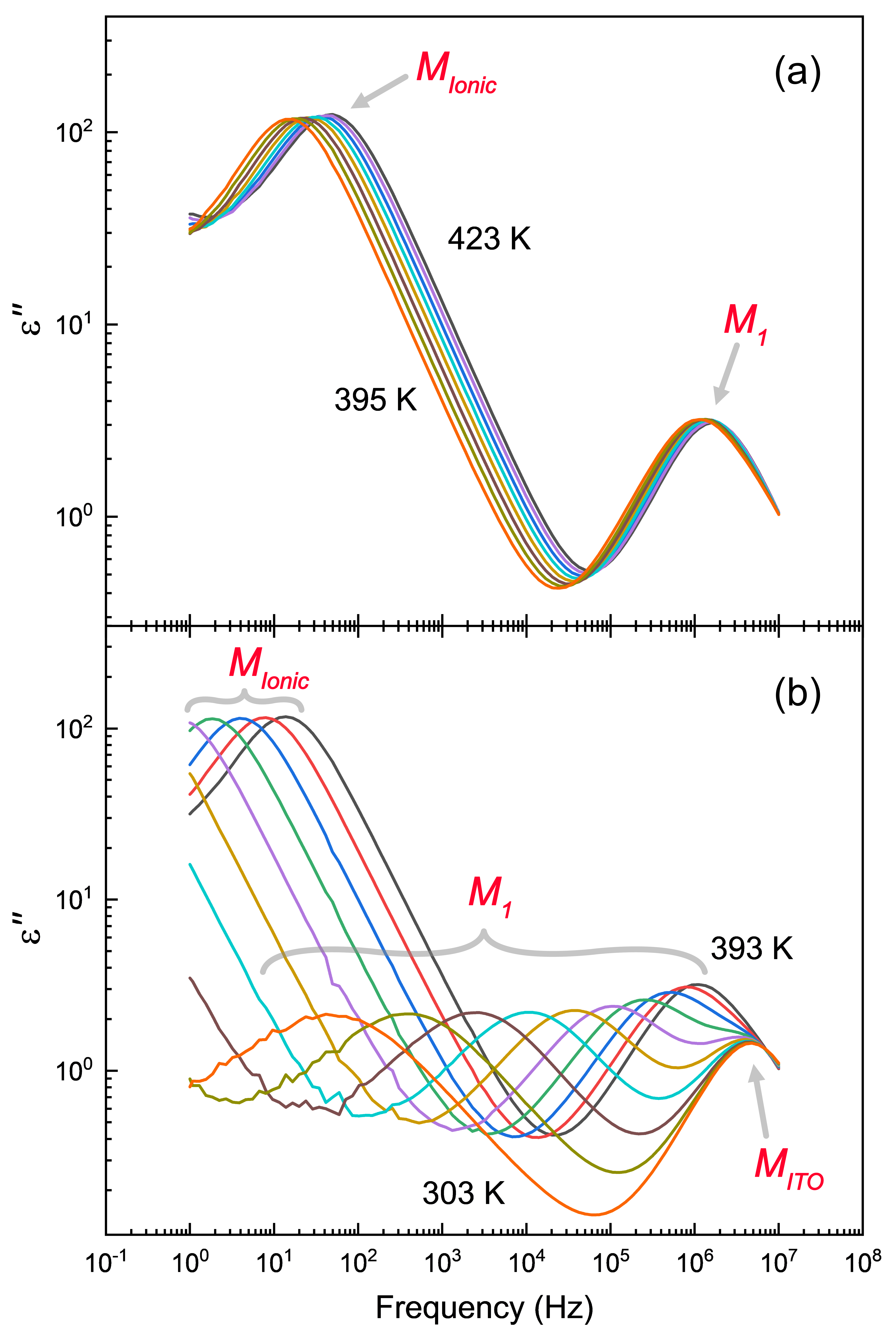}
    \caption{The variation of the imaginary part of the dielectric constant ($\epsilon^{\prime\prime}$) as a function of frequency at different temperatures. (a) In high temperature range from 423~K to 395~K with a temperature step of 4~K. (b) In the low temperature range from 393~K to 303~K with a temperature step of 10~K. The peaks in the data represent various dielectric relaxation processes.}
    \label{DieleRelax}
\end{figure}

The frequency dependence of $\epsilon^{\prime\prime}(f)$ at different temperatures in the dSmA and SmC phases are shown in figure~\ref{DieleRelax}. The observed peaks in $\epsilon^{\prime\prime}(f)$ curve correspond to the relaxation frequency of different dielectric modes of the sample. Figure~\ref{DieleRelax}a illustrates the frequency dependence of $\epsilon^{\prime\prime}(f)$ in the high temperature range from 423~K to 393~K with a temperature step of 4~K. The sample is in the dSmA phase at 423~K, and the transition to the SmC phase occurs at about 403~K. The relaxation peaks in the small frequency range below 100~Hz with relatively high dielectric strength are arising due to the ionic conductivity of the sample and will be ignored in further discussions. Another mode with the relaxation frequency in the MHz range was found in this temperature range, which is denoted as $M_1$ in the figure~\ref{DieleRelax}a. The intensity of these relaxation peaks remained almost constant, and the peak frequency gradually decreased upon lowering the temperature. The $M_1$ mode was found to exist in both dSmA and SmC phases and does not vary appreciably across the transition, as can be seen from figure~\ref{DieleRelax}a. The frequency dependence of $\epsilon^{\prime\prime}(f)$ in the lower temperature range from 393~K to 303~K with a temperature step of 10~K is shown in figure~\ref{DieleRelax}b. At lower temperatures, the relaxation frequency of mode $M_1$ started to decrease rapidly with decreasing temperature and attained a value of about 50~Hz at ambient temperature. In this lower temperature range, another weak relaxation peak at about 50~MHz was clearly visible, which did not vary with temperature as shown in figure~\ref{DieleRelax}b. This peak, denoted as $M_{ITO}$, arises due to the finite sheet resistance of the ITO coating used in the LC cell and is masked by the $M_1$ mode in the higher temperature range. 

Similar experiments were carried out on a homeotropically aligned sample, and one relaxation mode was found to exist in the same temperature range as in the planar aligned sample. The temperature variation of $\epsilon^\prime$ for the homeotropic and the planar aligned sample is shown in figure~S5 of SI. In the homeotropic LC cell, the molecules acquired a quasi-homeotropic alignment with a non-uniform dark texture between crossed polarizers, as shown in figure~S6 of SI. This non-uniform alignment is probably due to the incompatibility of the strong homeotropic anchoring and the diffuse cone structure of the dSmA phase. The quasi-homeotropic alignment of the molecules in the homeotropic cell can be attributed to the presence of mode $M_1$. Thus the sample exhibit one dielectric relaxation mode $M_1$ present in both alignments. The dielectric strength in the homeotropic aligned sample is significantly lower compared to that in the planar cells, indicating that the mesophases possess negative dielectric anisotropy.

In order to analyze the measured dielectric relaxation processes, the dielectric spectra were fitted using the Havriliak-Negami (HN) equation \cite{havriliak_1966, havriliak_1967}. This empirical equation expresses the frequency-dependent complex permittivity $\epsilon^{*}$ in terms of the various relaxation processes given by,
\begin{equation}
    \epsilon^{*}(f) - \epsilon_\infty = -\frac{i\sigma_0}{\epsilon_0 \omega^s}+\sum_{j=1}^{n} \frac{\Delta\epsilon_j}{\{1+(i\omega\tau_j)^{\alpha_j}\}^{\beta_j}}
    \label{HN_eqn}
\end{equation}
In this equation, $\Delta\epsilon_j$ represents the dielectric strength of the $j$th relaxation process, $\tau_j$ represents the corresponding relaxation time, $\epsilon_\infty$ represents the high-frequency limit of permittivity, and $\alpha_j$, $\beta_j$ are shape parameters. These parameters describe the broadness and asymmetry of the dielectric loss spectra, respectively, and satisfy the conditions $0<\alpha_j<1$ and $0<\alpha_j\beta_j<1$. The term $i\sigma_0/\epsilon_0 \omega^s$ is related to the conductivity, where $\sigma_0$ is the direct current (DC) conductivity, $\epsilon_0$ is the permittivity of free space and $s$ is a fitting parameter that determines the slope of the conductivity. In the case of pure Ohmic conductivity, $s=1$, while $s < 1$ could be observed in the case of additional influence due to electrode polarization \cite{Kremer_2003}. The HN response reduces to the Cole-Davidson response \cite{davidson_1951} when $\alpha = 1$, and to the Cole-Cole response \cite{cole_1941} when $\beta = 1$. The process under discussion is termed Debye relaxation for both $\alpha$ and $\beta$ equal to unity.
\begin{table}[b!]
\caption{\label{FitTable}The shape parameters $\alpha$ and $\beta$, and dielectric strength $\Delta\epsilon$ value obtained from fitting the experimental data with eqn.~\ref{HN_eqn} at different temperatures.}
\begin{ruledtabular}
\begin{tabular}{cccc}
T [K] & $\alpha$ & $\beta$ & $\Delta\epsilon$\\
\hline
303& 0.65 $\pm$ 0.02 & 1 $\pm$ 0.08 & 6.69 $\pm$ 0.17 \\
308& 0.68 $\pm$ 0.01 & 1 $\pm$ 0.06 & 6.50 $\pm$ 0.16  \\
313& 0.74 $\pm$ 0.02 & 0.95 $\pm$ 0.03 & 6.44 $\pm$ 0.21  \\
318& 0.67 $\pm$ 0.01 & 0.99 $\pm$ 0.06 & 7.33 $\pm$ 0.07  \\
323& 0.68 $\pm$ 0.01 & 1 $\pm$ 0.06 & 7.23 $\pm$ 0.06  \\
328& 0.67 $\pm$ 0.02 & 1 $\pm$ 0.08 & 7.23 $\pm$ 0.07  \\
333& 0.68 $\pm$ 0.02 & 1 $\pm$ 0.09 & 7.12 $\pm$ 0.08  \\
338& 0.69 $\pm$ 0.02 & 1 $\pm$ 0.12 & 7.06 $\pm$ 0.11  \\

\end{tabular}
\end{ruledtabular}
\end{table}

\begin{figure}[t]
    \centering
    \includegraphics[width=0.5\linewidth]{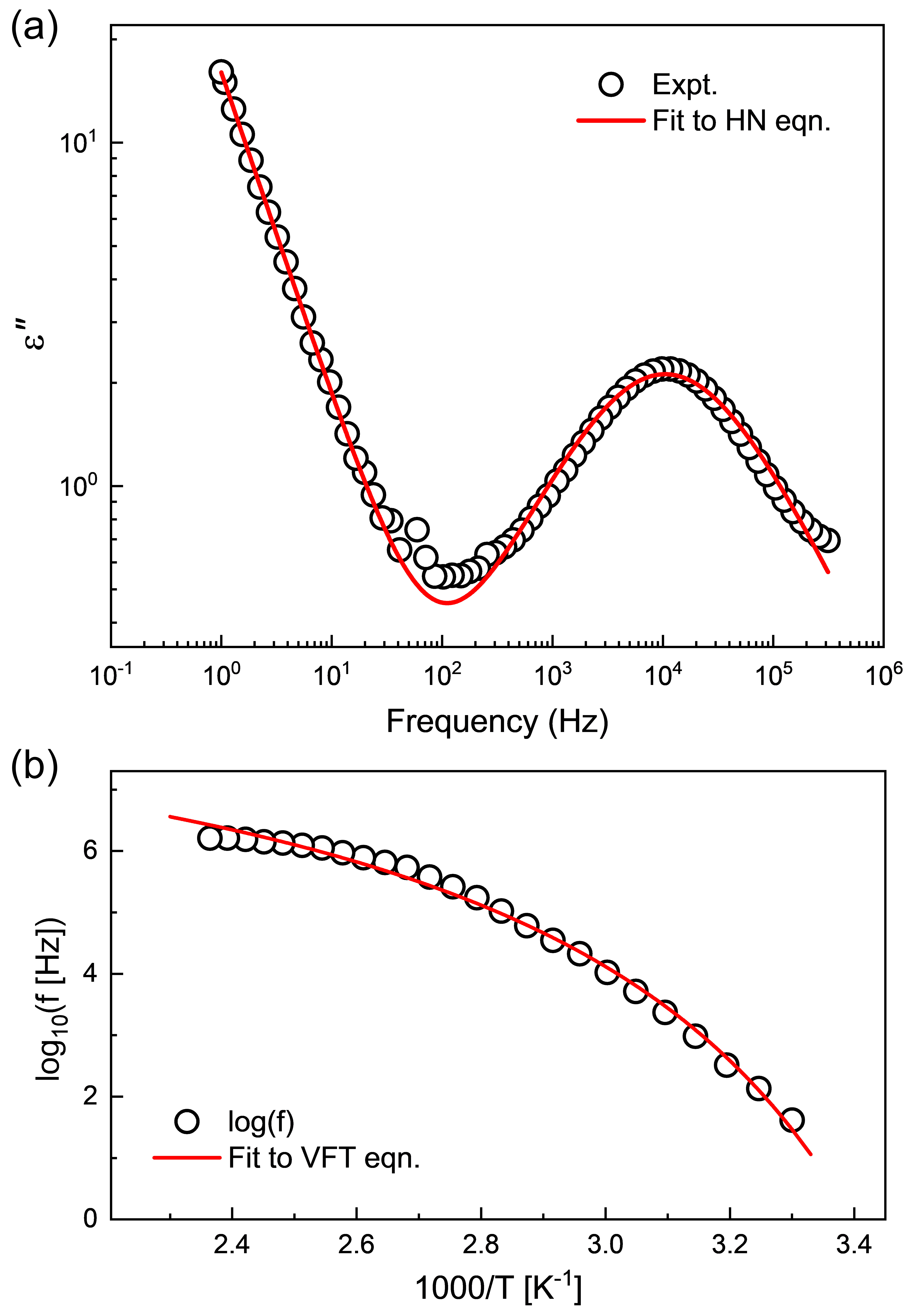}
    \caption{(a) Frequency variation of $\epsilon^{\prime\prime}$ at 333~K along with fitted curve (solid line) using eqn.~\ref{HN_eqn}. The deviation of the fitted curve in the higher frequency region is due to the overlap of the relaxation peak in the MHz range associated with ITO coating. (b) Relaxation frequency of mode $M_1$ as a function of inverse temperature. The solid curve represents the fit to the experimental data using VFT eqn.~\ref{VFT}, indicating the non-Arrhenius behavior of the liquid.} 
    \label{dieleFit}
\end{figure}

The dielectric spectra for our sample could be fitted well using eqn.~\ref{HN_eqn}. Figure~\ref{dieleFit}a shows the variation of $\epsilon^{\prime\prime}(f)$ as a function of frequency at 333~K along with the fitted curve. The fitting parameters obtained for some temperatures are listed in table~\ref{FitTable}. The asymmetry parameter $\beta = 1$ for our sample remains almost constant at varying temperatures. This implies that the shape of the dielectric loss peak is symmetric, and the relaxation process is of the Cole-Cole type. The Cole-Cole parameter $\alpha$ was found to be close to $0.68$ at lower temperatures and tends to decrease slightly with increasing temperature. This indicates that the width of the dielectric loss spectrum tends to increase with decreasing temperature, as can also be seen from figure~\ref{DieleRelax}b. The deviation of experimental data from the fitted curve at higher frequencies in figure~\ref{dieleFit}a is due to the overlap of the relaxation peak with the additional peak at about 50~MHz arising from the ITO coating of the LC cell.

The temperature dependence of relaxation frequency provides a useful classification of glassformers along a `strong' to `fragile' scale \cite{Angell_1995, Debenedetti_2001}. The former shows an Arrhenius dependence, whereas the latter deviates from the Arrhenius behavior. The temperature variation of the relaxation frequency for the $M_1$ mode of our compound can be fitted well with the modified Vogel-Fulcher-Tammann (VFT) equation \cite{Angell_1995} given by, 
\begin{equation}
    f = f_{\infty}\:\text{exp}\left(-\frac{DT_0}{T-T_0}\right)
    \label{VFT}
\end{equation}
where $f_\infty$ is the pre-exponential constant, $T_0$ is Vogel temperature, and $D$ is a constant that determines the deviation of the system away from the Arrhenius behavior, i.e., the fragility of the system. The measured relaxation frequency as a function of temperature along with its fit to eqn.~\ref{VFT} is shown in Figure~\ref{dieleFit}b. The fitting parameters are $T_0=253.15\pm0.004$~K, log$_{10}f_\infty=8.49\pm0.19$, and $D=3.19\pm0.39$. Such a large $D$ value indicates the highly fragile glassy nature of our sample. These fragile glasses have also been found for some polymeric systems \cite{Angell_1990}. The deviation of VFT fit from the experimental data in the higher temperature range can be attributed to the increasing overlap of the relaxation peak with the ITO-induced peak. It has also been reported that a poorer fit to the experimental data is expected for more fragile liquid \cite{Angell_1995}.
\begin{figure}[b]
    \centering
    \includegraphics[width=0.6\linewidth]{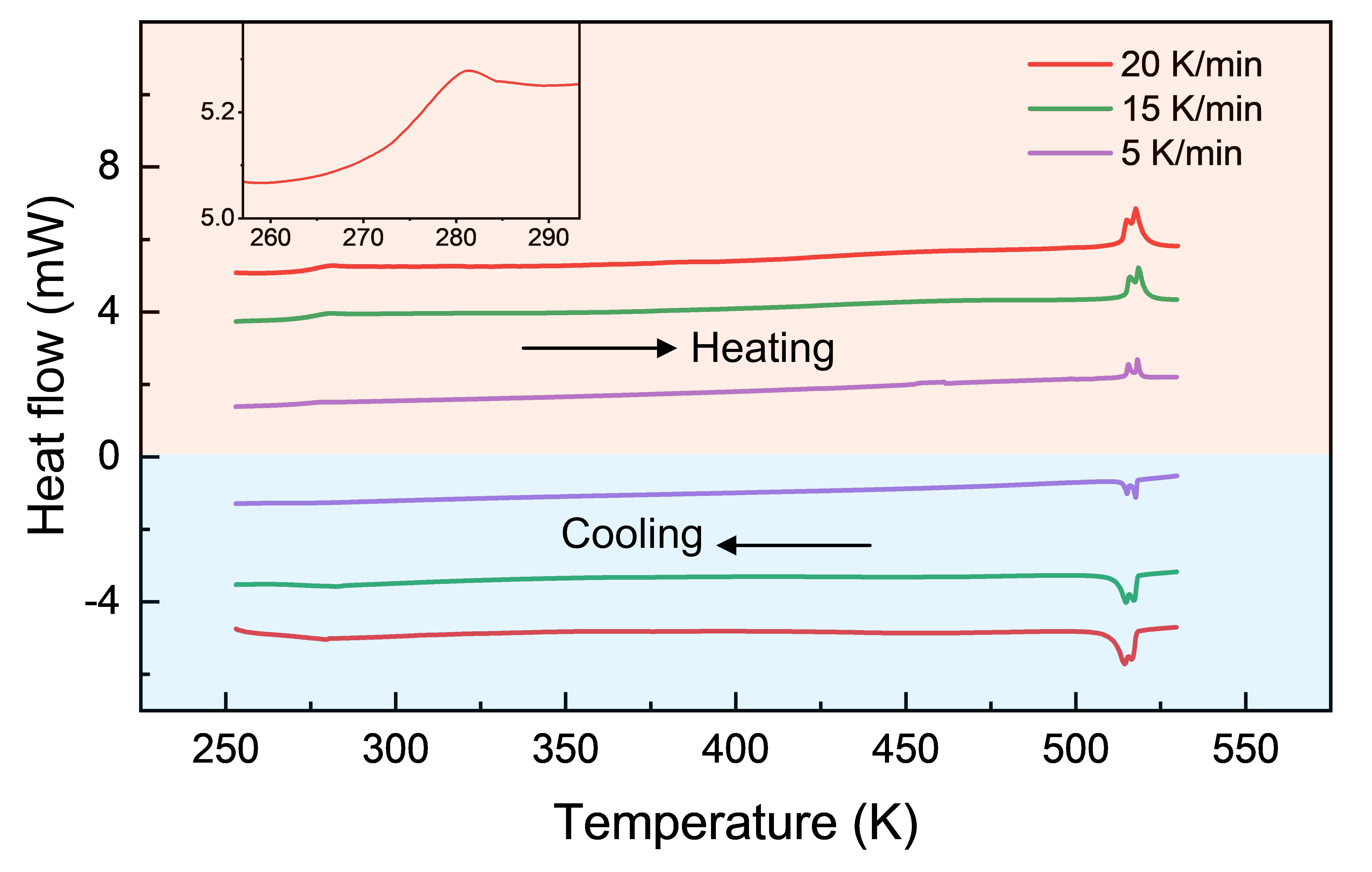}
    \caption{The DSC thermogram of the compound BTCN8 at different rates while cooling from the isotropic phase and subsequent heating. A step change in the curve at a temperature of 271~K is indicative of glass transition. The sample does not crystallize over a few weeks.}
    \label{dsc2}
\end{figure}

The DSC measurements were carried out to measure the possible glass transition in the cooling and heating cycle of the sample at different rates. The observed DSC thermograms are shown in figure~\ref{dsc2}. A step change in the DSC thermogram corresponding to the glass transition was detected on both the cooling and heating cycle. The vitrification temperature ($T_g$) on cooling at a rate of 20~K/min was 271~K, and the corresponding glass softening temperature on heating was 268~K. These temperatures were determined from the inflection point of the DSC thermogram corresponding to the half height of the step in the heat flow curve. The difference between calorimetric glass transition temperature $T_g$ and Vogel temperature $T_0$ is about 18~K for our sample, which is considered to be low as expected for highly fragile systems \cite{Angell_1995}. This difference tends to zero for magnetic relaxation in spin glasses \cite{Souletie_1990}. 

Now we discuss the possible origin of the observed $M_1$ mode in our sample. The dielectric relaxation processes can be divided broadly into two categories: non-collective and collective dynamics of the molecules. The liquid crystal molecules themselves can exhibit two dielectric relaxation mechanisms associated with the rotation of the molecules about their long and short axes. When the rotational dynamics of the individual molecules are not correlated, the resulting non-collective dielectric relaxation processes usually occur in the GHz or higher frequency range for typical rod-like molecules. The observed $M_1$ mode in our sample has a much lower relaxation frequency and is expected to be not associated with these non-collective rotational dynamics of the molecules. Thus, the $M_1$ mode must arise from the collective movement of the constituent molecules. Although the BTCN8 molecules have non-zero dipole moments, our experimental studies confirmed the absence of spontaneous polarization in the observed smectic phases. Furthermore, we didn't find experimental evidence of Langevin-type ordering of the dipole moments in the smectic layers \cite{Green_2019}. Hence, mode $M_1$ is not associated with the collective polar ordering of the molecules in our sample. 

\begin{figure}[t!]
    \centering
    \includegraphics[width=0.6\linewidth]{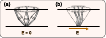}
    \caption{Schematic representation for the arrangement of bent-core molecules in the dSmA phase (a) without field, and (b) in the presence of the field. The reorientation of molecular distribution on field application gives rise to collective mode $M_1$ observed in our sample.}
    \label{model}
\end{figure}

Based on the molecular configuration in the dSmA and the SmC phases, we propose the following mechanism that gives rise to the $M_1$ mode. In the dSmA phase, the long axes of the molecules are distributed uniformly on a cone about the layer normal (see figure~\ref{model}a). Under the application of a small electric field in the plane of the smectic layers, the molecules reorient azimuthally on the cone resulting in a weak biaxiality in the system. Due to the strong transverse component of the dipole moment of the molecules, the sample has negative dielectric anisotropy, and the field tends to redistribute the molecules on the cone along the directions perpendicular to the applied field as shown in figure~\ref{model}b. The observed relaxation mode $M_1$ in our sample can be attributed to this collective redistribution of the molecules on the cone. In the SmC phase, the molecular distribution around the cone is peaked at a preferred azimuthal angle giving rise to a nonzero optical tilt. The optical tilt increases with decreasing temperature in the SmC phase. However, the molecules still have a broad distribution along the azimuthal direction. This is also supported by the fact that the optical tilt angle in the SmC phase is about 15$^\circ$ at ambient temperature, which is significantly smaller than the corresponding intralayer molecular tilt measured from XRD studies ($\sim$ 33$^\circ$). Hence, a similar molecular redistribution confining them in the plane perpendicular to the applied field is expected to occur in the planar aligned SmC phase. Thus, the relaxation frequency of the mode $M_1$ varies continuously across the transition from the dSmA to the SmC phase and persists throughout the SmC phase. The relaxation frequency of mode $M_1$ decreases gradually with decreasing temperature, indicating that this molecular redistribution about the cone tends to be frozen. The relaxation frequency starts to decline rapidly in the lower temperature region of the SmC phase because of the experimentally confirmed glassy behavior of our sample.

The broad molecular distribution on the tilt cone also explains the increase in birefringence of the SmC phase under the application of a relatively higher field of about 2~V/$\mu$m, as observed in POM studies (see figure~S3 of SI). The applied field confines the molecules in the plane of the LC cell. This results in increased order parameters and hence the birefringence of the sample. This field induced change is reversible as the molecules relax back to their original distribution after removal of the field. 

The constituent molecules of the compound BTCN8 have a central thiophene ring which gives rise to an opening angle of about 143$^\circ$, which is larger than 120$^\circ$ typically found for bent-core molecules. This perhaps can be attributed to the absence of net polarization in the observed smectic phases. We do not see any sign of the B2 phase typically exhibited by bent-core molecules. Rather, the observed phases are calamatic phases generally observed for rod-like molecules.

\section{Conclusion} 
We report the experimental studies on a compound consisting of bent-core banana shaped molecules, which exhibits enantiotropic liquid crystalline phases following the sequence \emph{Isotropic $\rightarrow$ Nematic $\rightarrow$ dSmA $\rightarrow$ SmC} on cooling. The dSmA to the SmC transition is accompanied by minimal layer contraction. The layers do not possess spontaneous polarization in both the smectic phases. A dielectric relaxation mode was observed due to the reorientation of the molecules on the tilt cone in the dSmA phase and it persists in the lower temperature SmC phase. The relaxation frequency decreases rapidly following the VFT equation indicating a fragile glassy behavior of our sample. The DSC measurements confirmed the glass transition at a temperature of about 271~K, which do not have significant dependence on the cooling rate. We have proposed a model for the observed dielectric mode in the dSmA phase to account for experimental results. Interestingly, the studied bent-core compound has the unique property of exhibiting a de Vries SmA phase along with the glassy behavior. 

\begin{acknowledgments}
We thank Ms. Vasudha K. N. for her help in acquiring DSC and XRD data. We also thank Dr. T. Narasimhaswamy for his support in the chemical synthesis of the compound.  
\end{acknowledgments}
%

\appendix
\renewcommand\thefigure{S\arabic{figure}} 

\section{Optical tilt}
The optical tilt $\theta_{opt}$ in the SmC phase of compound BTCN8 is the angle between the optic axis and the layer normal. In the SmC phase, two oppositely tilted domains were observed for a planar aligned sample of thickness 5~$\mu$m in POM between crossed polarizers. These domains appear optically identical when the rubbing direction of the LC cell is parallel, i.e., $0^\circ$ to the analyzer as shown in figure~\ref{pom2}b. On rotating the sample stage clockwise (anticlockwise), $\text{domain~\RNum{1}}$ ($\text{domain~\RNum{2}}$) becomes darkest when the rubbing direction makes an angle of about $\mp$11$^\circ$ with respect to the analyzer as shown in figure~\ref{pom2}a (\ref{pom2}c). A domain appears darkest when the local optic axis aligns with the analyzer, and the corresponding rotation angle gives the optical tilt angle $\theta_{opt}$. Figure~\ref{pom2}d shows the detailed variation of normalized intensity as a function of the rotation angle in domains $\text{\RNum{1}}$ and $\text{\RNum{2}}$. The two domains appear optically identical at angles -90$^\circ$, -45$^\circ$, 0$^\circ$, 45$^\circ$, and 90$^\circ$ due to the symmetric optical tilt in opposite directions in the two domains. The minima of the curves near $0^\circ$ give the $\theta_{opt}$, and the brightest positions occur at an angle of 45$^\circ$ from the darkest positions in the respective domains.  
\setcounter{figure}{0} 
\begin{figure}[h]
    \centering
    \includegraphics[width=0.6\textwidth]{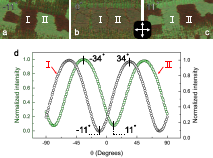}
    \caption{The POM textures of a planar aligned sample of compound BTCN8 in the SmC phase at 373~K for three orientations (a) $-11^\circ$, (b) 0$^\circ$, and (c) $11^\circ$ of rubbing direction of the sample cell with respect to the analyzer. Two oppositely tilted domains indicated by $\text{\RNum{1}}$ and $\text{\RNum{2}}$ were observed which appears identical at 0$^\circ$. The brightness of the domains $\text{\RNum{1}}$ and $\text{\RNum{2}}$ interchange on rotating the sample clockwise and counterclockwise. (d) The variation of the normalized intensity of the $\text{domain~\RNum{1}}$ and $\text{domain~\RNum{2}}$ as a function of the rotation angle $\theta$ of the rubbing direction of the LC cell with respect to the analyzer. The minima at $\mp$11$^\circ$ corresponds to the optical tilt angle in the domain $\text{\RNum{1}}$ and $\text{\RNum{2}}$, respectively.}
    \label{pom2}
\end{figure}
\newpage
\section{Polarization current measurements}
\begin{figure}[h]
    \centering
    \includegraphics[width=0.8\textwidth]{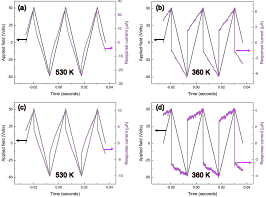}
    \caption{Time evolution of current response when the sample was subjected to a triangular AC voltage of amplitude 50~V and frequency 40~Hz for planar alignment in (a) dSmA and (b) SmC phase, and homeotropic alignment in (c) dSmA and (d) SmC phase, respectively. The absence of a current peak due to polarization reversal indicates that both the smectic phases are apolar in nature.}
    \label{scr}
\end{figure}

The triangular wave technique \cite{Miyasato_1983} was utilized in order to determine the state of spontaneous polarization in the de Vries SmA (dSmA) and SmC phases of our compound. A triangular wave voltage of amplitude 50~V at frequency 40~Hz was applied across the sample. The current response of a 5~$\mu$m thick planar-aligned sample in the dSmA and SmC phase are shown in figure~\ref{scr}a and \ref{scr}b, respectively. The absence of peaks corresponding to polarization reversal current in both phases confirms that the layers do not possess spontaneous polarization. Similar experiments were performed on a 9~$\mu$m thick homeotropically aligned sample. The current response for this alignment in the dSmA and SmC phase are shown in figure~\ref{scr}c and \ref{scr}d, respectively. No polarization reversal current was observed again, confirming the apolar nature of the layers.
\newpage
\section{Electro-optical measurements}
\begin{figure}[h]
    \centering
    \includegraphics[width=0.8\textwidth]{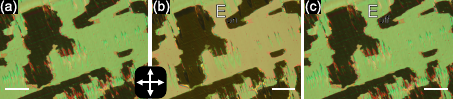}
    \caption{The field induced changes in the POM textures of a planar aligned sample in the SmC phase at 360~K between crossed polarizers. The texture (a) at initial zero fields, (b) with a triangular field of amplitude 2~V/$\mu$m and frequency~f, and (c) after removing the field. The length of the bar shows a scale of 50~$\mu$m.}
    \label{pom3}
\end{figure}

The textural changes in the dSmA and SmC phases were observed with the application of a triangular wave voltage. The texture in the dSmA phase does not change appreciably with the applied field, even at an amplitude of about 10~V/$\mu$m. On the other hand, the color of the texture changes slightly in the lower temperature SmC phase even at a small field of about 2~V/$\mu$m as shown in figure~\ref{pom3}a and \ref{pom3}b. The changes in the color indicate a slight increase in the birefringence of the sample, which arises due to more localized molecular distribution on the tilt cone, as described in the main text. This field-induced transformation is reversible in nature as the original texture can be retrieved after removing the field, as shown in figure~\ref{pom3}c. 

\begin{figure}[h]
    \centering
    \includegraphics[width=0.6\textwidth]{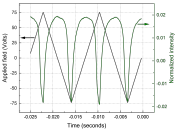}
    \caption{The normalized optical transmittance along with the applied AC voltage as a function of time through a planar aligned sample of thickness 5~$\mu$m kept between crossed polarizers at 393~K.}
    \label{pom4}
\end{figure}

The optical transmittance of the sample in the SmC phase was measured as a function of time under the application of a triangular wave voltage of amplitude 75~V and frequency 80~Hz. The details of the experimental setup are given in the main text. The steady-state normalized optical transmittance through a planar aligned sample kept between crossed polarizers at 393~K is shown in figure~\ref{pom4}. The figure~\ref{pom4} also depicts the time trace of the applied triangular wave voltage. The optical signal has a 2$f$ response with respect to the applied electric field, suggesting a quadratic coupling between the field and the dielectric anisotropy of the sample.  This result also indicates the absence of polarization in the SmC phase. As seen from figure~\ref{pom4}, the transmitted intensity is maximum at zero fields and decreases with increasing magnitude of the field. This is due to the optical phase difference ($\Delta\phi$) lying between $\pi$ and $2\pi$ for our 5~$\mu$m thick sample.
\section{Dielectric studies}
\begin{figure}[h]
    \centering
    \includegraphics[width=0.90\textwidth]{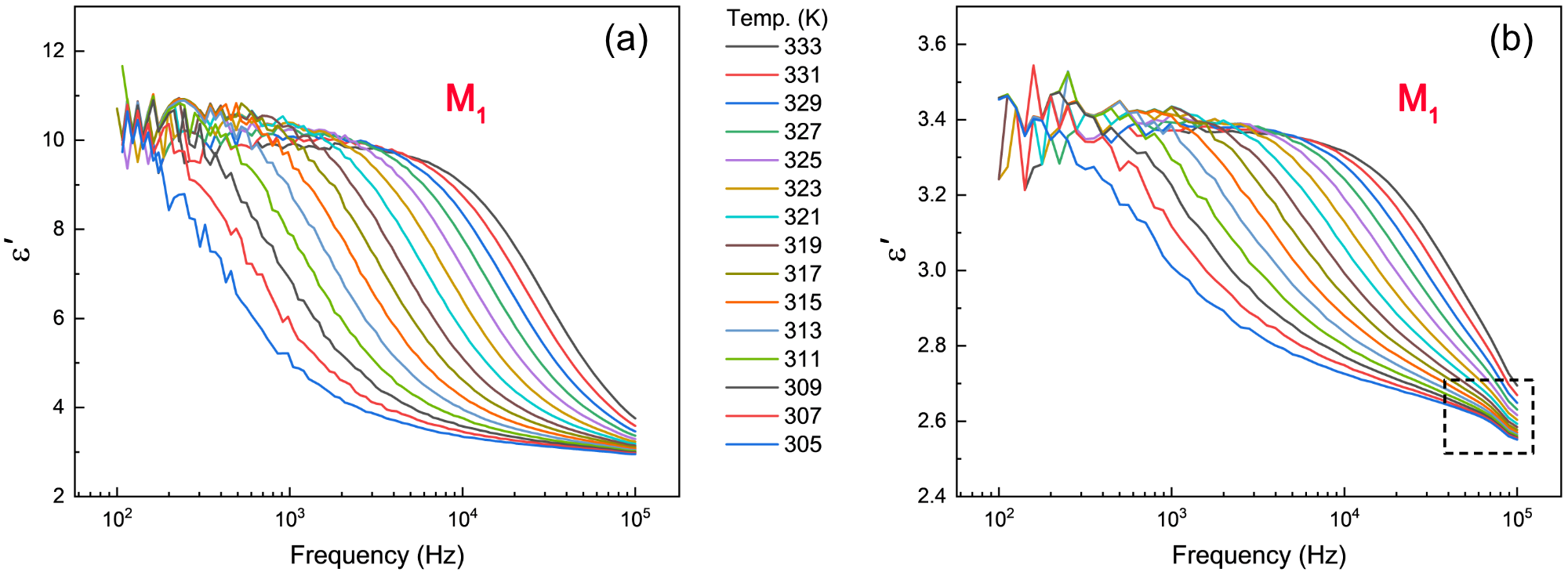}
    \caption{Frequency dependence of the real part of dielectric constant $\epsilon^\prime$ for (a) planar aligned sample and (b) homeotropically aligned sample at different temperatures. The noisy data in the lower frequency domain is due to experimental artifacts. The temperature depicted in the legend applies to both graphs. A slight change in the data enclosed by the dotted rectangle is not a sample property. Rather, it occurs due to the LC cell as it was found to exist in the measurements with empty cells too.}
    \label{dielectric2}
\end{figure}

\begin{figure}[h]
    \centering
    \includegraphics[width=0.65\textwidth]{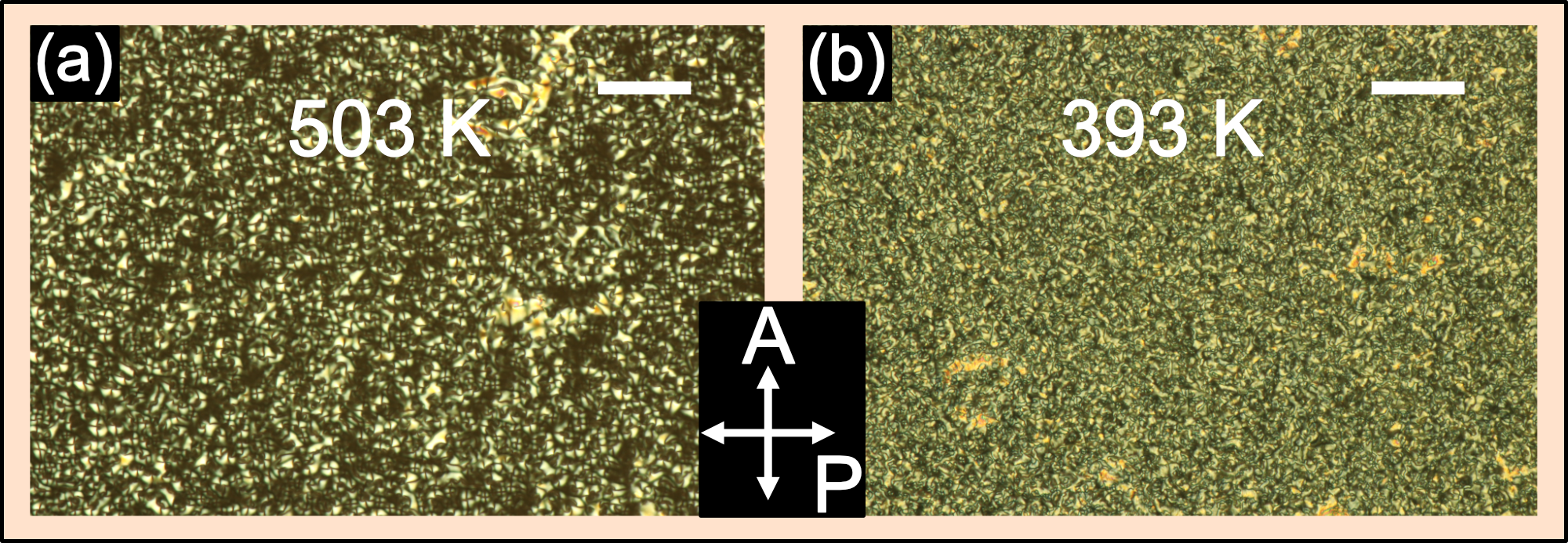}
    \caption{The POM texture of the sample in a homeotropic LC cell of thickness 9~$\mu$m under crossed polarizers in (a) the dSmA phase, and (b) the SmC phase. The scale bar indicates 100~$\mu$m.}
    \label{Hcell_POM}
\end{figure}

The real part of the dielectric constant was also measured in a homemade custom setup as a function of frequency for planar as well as homeotropic aligned samples at various temperatures. For the planar aligned sample, the dielectric relaxation shown in figure~\ref{dielectric2}a agrees very well with the measurements performed using a commercially available setup (Novocontrol Alpha~A). This further confirms that the relaxation process arises from the sample. In commercially available homeotropic LC cells of thickness 9~$\mu$m, the molecular alignment is found to be quasi-homeotropic with a large number of bright spots in the POM texture between crossed polarizers, as shown in figure~\ref{Hcell_POM}. The dielectric measurements in this homeotropic cell also show the presence of the relaxation mode $M_1$ with much reduced dielectric strength (see figure~\ref{dielectric2}b). This clearly indicates the negative dielectric anisotropy of the sample in the observed smectic phases. However, the presence of the mode $M_1$ in the homeotropic sample is not expected; it perhaps arises due to the quasi-homeotropic nature of the molecular alignment in the cell. 

The POM textures of the sample in a homeotropic cell of thickness 9~$\mu$m under crossed polarizers are shown in figure~\ref{Hcell_POM}. While cooling the sample to the dSmA phase, the molecules tend to align homeotropically, having a dark texture interspersed with numerous bright spots, as shown in figure~\ref{Hcell_POM}a. This texture indicates a quasi-homeotropic alignment of the molecules. In the SmC phase at lower temperatures, the sample acquires an inhomogeneous grainy texture, as shown in figure~\ref{Hcell_POM}b, instead of an expected schlieren texture.    

\end{document}